\documentstyle[12pt]{article}
\def\Journal#1#2#3#4{{#1} {\bf #2}, #3 (#4)}

\def\NCA{\em Nuovo Cimento}
\def\CR{\em C.R. Acad. Sci. (Paris)}
\def\CP{\em Cahiers de Physique}
\def\PRL{\em Phys. Rev. Lett.}
\def\JMP{\em J. Math. Phys.}
\def\GRG{\em Gen. Rel. Grav.}
\def\CQG{\em Class. Quantum Grav.}

\def\be{\begin{equation}}
\def\ee{\end{equation}}
\def\bea{\begin{eqnarray}}
\def\eea{\end{eqnarray}}
\def\ben{\begin{eqnarray*}}
\def\een{\end{eqnarray*}}

\newcommand{\bm}[1]{\mbox{\boldmath $#1$}}

\begin{document}

\title{(Super)$^n$-Energy for Arbitrary Fields and its Interchange:
Conserved Quantities\footnote{This essay received an ``honorable
                        mention" in the 1999 Essay Competition of the
                        Gravity Research Foundation.}}

\author{Jos\'{e} M.M. Senovilla\\
Departament de F\'{\i}sica Fonamental, Universitat de Barcelona,\\
Diagonal 647, 08028 Barcelona, Spain.\\
e-mail: seno@ffn.ub.es}

\date{March 15, 1999}
\maketitle

\thispagestyle{empty}
\abstract{Inspired by classical work of Bel and Robinson, a natural purely
algebraic construction of super-energy (s-e) tensors for {\it arbitrary} fields
is presented, having good {\it mathematical} and {\it physical} properties.

Remarkably, there appear quantities with mathematical characteristics of energy
densities satisfying the {\it dominant property}, which provides s-e 
estimates useful for global results and helpful in other matters.

For physical fields, higher order (super)$^n$-energy tensors involving the
field and its derivatives arise. In Special Relativity, they provide infinitely
many conserved quantities.

The {\it interchange} of s-e between different fields is shown. The
discontinuity propagation law in Einstein-Maxwell fields is related to s-e
tensors, providing quantities conserved along null hypersurfaces. Finally,
conserved s-e currents are found for any minimally coupled scalar field
whenever there is a Killing vector. 
}
\newpage

The importance of the Bel-Robinson and other super-energy (s-e) tensors
\cite{B1,B4} is usually recognized, even though their physical
interpretation remains somewhat obscure. Their {\it mathematical} usefulness is
clear, a manifestation of which is their relevance in the proof of the
global stability of Minkowski spacetime \cite{CK}, or in the study of well-posed
symmetric hyperbolic systems of differential equations including gravity
(see, e.g. \cite{F}). In this paper, I show how to generalize these
important mathematical properties to arbitrary fields and, more importantly,
try to shed some light into the physical meaning of s-e by finding non-trivial
conservation laws which involve {\it two} different fields.

First, let us present the procedure to construct the s-e of any given tensor
\cite{S}. Consider any tensor $t_{\mu_1\dots\mu_m}$ as an
{\it r-fold $(n_1,\dots,n_r)$-form} ($n_1+\dots +n_r =m$) by separating the
$m$ indices into $r$ blocks, each containing $n_A$ ($A=1,\dots,r$) completely
antisymmetric indices. This can always be done. Several examples are:
$F_{\mu\nu}=F_{[\mu\nu]}$ is a simple (2)-form, while $\nabla_{\rho}F_{\mu\nu}$
is a double (1,2)-form; the Riemann tensor $R_{\alpha\beta\lambda\mu}$ is a
double {\it symmetrical} (2,2)-form (pairs can be interchanged); the
Ricci tensor $R_{\mu\nu}$ is a double symmetrical (1,1)-form, etcetera.
Schematically, $t_{\mu_1\dots\mu_m}$ will be denoted
by $t_{[n_1],\dots,[n_r]}$ where $[n_A]$ indicates the $A$-th block with $n_A$
antisymmetrical indices.
Duals can be defined by contracting each of the blocks with the volume element
4-form, obtaining the tensors (obvious notation, signature --,+,+,+):
\ben
t_{\stackrel{*}{[4-n_1]},\dots,[n_r]}, \dots ,\hspace{2mm}
t_{[n_1],\dots,\stackrel{*}{[4-n_r]}},\hspace{2mm}
t_{\stackrel{*}{[4-n_1]},\stackrel{*}{[4-n_2]},\dots,[n_r]}, \, \dots ,
\hspace{2mm} t_{\stackrel{*}{[4-n_1]},\dots,\stackrel{*}{[4-n_r]}} \, .
\een
There are
$1+ \left(\begin{array}{c} r\\ 1\end{array}\right)+
\dots + \left(\begin{array}{c} r\\ r\end{array}\right)=2^r$
tensors in this set (including $t_{[n_1],\dots,[n_r]}$).

Let us define the ``semi-square''
$(t_{[n_1],\dots,[n_r]}\times t_{[n_1],\dots,[n_r]})$ by contracting all
indices but one of each block in the product of $t$ with itself
(after reordering indices if necessary)
\ben
(t\times t)_{\lambda_1\mu_1\dots\lambda_r\mu_r}\equiv
\left(\prod_{A=1}^{r}\frac{1}{(n_A-1)!}\right)
t_{\lambda_1\rho_2\dots\rho_{n_1},\dots ,\lambda_r\sigma_2\dots\sigma_{n_r}}
t_{\mu_1\hspace{10mm}\dots ,\mu_r}^{\hspace{2mm}\rho_2\dots\rho_{n_1}
,\hspace{5mm}\sigma_2\dots\sigma_{n_r}} \, .
\een
The {\it s-e tensor of $t$} is $2r$-covariant and defined as half of the
sum of the $2^r$ semi-squares constructed with $t_{[n_1],\dots,[n_r]}$ and all
its duals. Explicitly:
\bea
2\, T_{\lambda_1\mu_1\dots\lambda_r\mu_r}\left\{t\right\}\equiv
\left(t_{[n_1],\dots,[n_r]}\times t_{[n_1],\dots,
[n_r]}\right)_{\lambda_1\mu_1\dots\lambda_r\mu_r}
+\dots \, \, \dots \, +\nonumber \\ 
\left(t_{\stackrel{*}{[4-n_1]},\dots,\stackrel{*}{[4-n_r]}}\times
t_{\stackrel{*}{[4-n_1]},\dots,
\stackrel{*}{[4-n_r]}}\right)_{\lambda_1\mu_1\dots\lambda_r\mu_r} \, .
\label{set}
\eea
It can be proven that (\ref{set}) is symmetric on each
$(\lambda_A\mu_A)$-pair; if $t_{[n_1],\dots,[n_r]}$ is symmetric in the
change of two $[n_A]$-blocks, then (\ref{set}) is symmetric with respect to
the corresponding $(\lambda_A\mu_A)$-pairs; and (\ref{set}) is traceless
in any $(\lambda_A\mu_A)$-pair with $n_A=2$.

For any timelike unit vector ${\vec u}$, the {\it s-e density of $t$ relative to}
${\vec u}$ is defined as 
\ben
W_t\left(\vec{u}\right)\equiv
T_{\lambda_1\mu_1\dots\lambda_r\mu_r}\{t\}
u^{\lambda_1}u^{\mu_1}\dots u^{\lambda_r}u^{\mu_r} \, .
\een
$W_t\left(\vec{u}\right)$ is non-negative and satisfies
\be
\left\{\exists \vec{u}\hspace{3mm} \mbox{such that} \hspace{2mm}
W_t\left(\vec{u}\right)=0 \right\} \Longleftrightarrow
T_{\lambda_1\mu_1\dots\lambda_r\mu_r}\{t\}=0 \Longleftrightarrow
t_{\mu_1\dots\mu_m}=0 .\label{cero}
\ee
Actually, $W_t\left(\vec{u}\right)$ is the sum of the squares
$|t_{\mu_1\dots\mu_m}|^2$ of all the components of $t$ in any orthonormal basis
$\{{\vec e}_{\mu}\}$ with $\vec{u}=\vec{e}_0$. More importantly, the
tensor (\ref{set}) satisfies the {\it dominant s-e property}
(DSEP) \cite{PR,BS,Ber,Ber2}, that is, for any future-pointing causal vectors
${\vec k}_1,\dots,{\vec k}_{2r}$ we have
\ben
T_{\lambda_1\mu_1\dots\lambda_r\mu_r}\{t\}
k^{\lambda_1}_1k^{\mu_1}_2\dots k^{\lambda_r}_{2r-1}k^{\mu_r}_{2r}\geq 0 \, .
\een
This is equivalent to the ``dominance'' of $T_{0\dots 0}\{t\}$ in any
orthonormal basis:
\ben
W_t\left(\vec{e}_0\right)=T_{0\dots 0}\{t\}\geq
\left|T_{\lambda_1\mu_1\dots\lambda_r\mu_r}\{t\}\right|, \hspace{3mm}
\forall \lambda_1, \mu_1, \dots, \lambda_r, \mu_r \, .
\een
The proof of DSEP, which is one of the main properties of definition
(\ref{set}), has been given by Bergqvist \cite{Ber2} in full generality
using spinors. DSEP is a helpful tool to prove causal propagation of
$t_{\mu_1\dots\mu_m}$ (see e.g.\ \cite{HE,BS,F})
and provides s-e estimates which may be useful in generalizing the
theorems in \cite{CK} to the case with matter.

Applying (\ref{set}) to the gravitational field, one constructs
the tensors $T^{\alpha\beta\lambda\mu}\left\{R_{[2][2]}\right\}$
and $T^{\alpha\beta\lambda\mu}\left\{C_{[2][2]}\right\}$ associated to the
Riemann and Weyl curvatures, getting the classical Bel \cite{B4} and
Bel-Robinson \cite{B1} (BR) tensors, respectively
\bea
2\, B^{\alpha\beta\lambda\mu}\equiv  \! R^{\alpha\rho\lambda\sigma}  
R^{\beta\hspace{1mm}\mu}_{\hspace{1mm}\rho\hspace{2mm}\sigma}\! +   
\! {*R}^{\alpha\rho\lambda\sigma}
{*R}^{\beta\hspace{1mm}\mu}_{\hspace{1mm}\rho\hspace{2mm}\sigma}\! +
\! {R*}^{\alpha\rho\lambda\sigma}
R*^{\beta\hspace{1mm}\mu}_{\hspace{1mm}\rho\hspace{2mm}\sigma}\! +   
{*R*}^{\alpha\rho\lambda\sigma}
{*R*}^{\beta\hspace{1mm}\mu}_{\hspace{1mm}\rho\hspace{2mm}\sigma}\, ,
\label{bel} \\
{\cal T}^{\alpha\beta\lambda\mu}\equiv
C^{\alpha\rho\lambda\sigma}
C^{\beta\hspace{1mm}\mu}_{\hspace{1mm}\rho\hspace{2mm}\sigma}+
\stackrel{*}{C}\hspace{.1mm}^{\alpha\rho\lambda\sigma}
\stackrel{*}{C}\hspace{.1mm}^{\beta\hspace{1mm}\mu}
_{\hspace{1mm}\rho\hspace{2mm}\sigma} \hspace{2cm} \nonumber\\
B^{\alpha\beta\lambda\mu}=B^{(\alpha\beta)(\lambda\mu)}=
B^{\lambda\mu\alpha\beta},\, \,      
B_{\hspace{1.5mm}\alpha}^{\alpha\hspace{1mm}\lambda\mu}=0, \hspace{7mm}
{\cal T}^{\alpha\beta\lambda\mu}={\cal T}^{(\alpha\beta\lambda\mu)}, \, \,
{\cal T}_{\hspace{2mm}\alpha}^{\alpha\hspace{1mm}\lambda\mu} =0 .\nonumber
\eea
These tensors have physical dimensions of $L^{-4}$, leading to two
possible interpretations: either as energy density ``square'' (\cite{BS2} and
references therein), or as energy density per unit surface. This second
interpretation seems correct, as can be deduced from several independent
results \cite{quasi,other}. Notably, the relationship \cite{quasi}
\ben
E_r = \mbox{const.\,} W_{{\cal T}}\left(\vec{u}\right)\mid_{r=0} r^5 + O(r^6)
\een
between the quasilocal energy $E_r$ of small 2-spheres of radius $r$
in vacuum and the BR s-e density $W_{{\cal T}}\left(\vec{u}\right)$ supports
clearly this view (${\vec u}$ orthogonal to the 2-sphere).

The Bel tensor satisfies
\be
\nabla_{\alpha}B^{\alpha\beta\lambda\mu}=
R^{\beta\hspace{1mm}\lambda}_{\hspace{1mm}\rho\hspace{2mm}\sigma}
J^{\mu\sigma\rho}+R^{\beta\hspace{1mm}\mu}_{\hspace{1mm}\rho\hspace{2mm}\sigma}
J^{\lambda\sigma\rho}-\frac{1}{2}g^{\lambda\mu}
R^{\beta}_{\hspace{1mm}\rho\sigma\gamma}J^{\sigma\gamma\rho} \label{div}
\ee
where $J_{\lambda\mu\beta}\equiv
\nabla_{\lambda}R_{\mu\beta}-\nabla_{\mu}R_{\lambda\beta}$. Therefore, the
Bel (as well as the BR) tensor is divergence-free in Einstein spaces (including
vacuum):
\ben
R_{\alpha\beta}=\Lambda g_{\alpha\beta} \hspace{1cm} \Longrightarrow \hspace{1cm}
\nabla_{\alpha}B^{\alpha\beta\lambda\mu}=
\nabla_{\alpha}{\cal T}^{\alpha\beta\lambda\mu}=0 
\een
implying that ${\cal T}^{\alpha\beta\lambda\mu}\xi_{\beta}\eta_{\lambda}
\zeta_{\mu}$ is divergence-free for any conformal Killing vectors ${\vec \xi}$,
${\vec \eta}$, ${\vec \zeta}$. Other divergence-free tensors appear
in \cite{SaCo}, but they do not satisfy DSEP nor (\ref{cero}), and their
lack of index symmetries prevents the existence of conserved currents.

Let us now apply our general procedure to physical fields. Take any 
electromagnetic field $F_{\mu\nu}$ satisfying the source-free Maxwell equations:
$\nabla_{\rho}F^{\rho}_{\hspace{-1mm}\nu}=0$, $\nabla_{[\rho}F_{\mu\nu]}=0$.
Using (\ref{set}) one arrives at
\ben
T_{\lambda\mu}\{F_{[2]}\}=F_{\lambda\rho}F^{\rho}_{\hspace{-1mm}\mu}-
\frac{1}{4}g_{\lambda\mu}F_{\rho\sigma}F^{\rho\sigma}
\een
which is the standard divergence-free energy-momentum tensor, leading to
conserved currents $T^{\mu\nu}\{F_{[2]}\}\zeta_{\nu}$ for any conformal Killing
vector ${\vec \zeta}$.  What about BR-like tensors? As a next step
one can use the double (1,2)-form $\nabla_{\alpha}F_{\mu\nu}$
as starting object to construct the corresponding tensor (\ref{set}),
which becomes
\bea
T_{\alpha\beta\lambda\mu}\{\nabla_{[1]} F_{[2]}\}=
\nabla_{\alpha}F_{\lambda\rho}\nabla_{\beta}F_{\hspace{-1mm}\mu}^{\rho}+
\nabla_{\alpha}F_{\mu\rho}\nabla_{\beta}F_{\hspace{-1mm}\lambda}^{\rho}-
g_{\alpha\beta}\nabla_{\sigma}F_{\lambda\rho}
\nabla^{\sigma}F_{\hspace{-1mm}\mu}^{\rho}- \nonumber \\
-\frac{1}{2}g_{\lambda\mu}
\nabla_{\alpha}F_{\sigma\rho}\nabla_{\beta}F^{\sigma\rho}+
\frac{1}{4}g_{\alpha\beta}g_{\lambda\mu}
\nabla_{\tau}F_{\sigma\rho}\nabla^{\tau}F^{\sigma\rho} \, .
\label{em}
\eea
Chevreton's tensor \cite{C} is
given simply by $T_{\alpha\beta\lambda\mu}\{\nabla_{[1]} F_{[2]}\}+
T_{\lambda\mu\alpha\beta}\{\nabla_{[1]} F_{[2]}\}$. The procedure can be
continued building the (super)$^2$-energy tensor
$T_{\alpha\beta\lambda\mu\tau\nu}$ associated with the triple (1,1,2)-form
$\nabla_{\alpha}\nabla_{\beta}F_{\lambda\mu}$, and
so on, producing an infinite set of (super)$^n$-energy tensors, one
for each natural number $n$. The following fundamental result holds:
{\it the (super)$^n$-energy (tensor) of the
electromagnetic field vanishes at a point $p$ iff the n$^{th}$ covariant
derivative of $F$ is zero at $p$}.

The divergences of (\ref{em}) and higher order s-e tensors can be
computed, and they {\it vanish in flat spacetime} providing infinitely many
conserved quantities for the electromagnetic field in Special Relativity.
This result is not surprising, because if $F_{\mu\nu}$ satisfies Maxwell's
equations in flat spacetime, then the 2-forms
$\nabla_{|\alpha_1|}\dots\nabla_{|\alpha_s|}F_{\mu\nu}$ for {\it fixed} values
of $\alpha_1,\dots,\alpha_s$ so do, and their
``energy-momentum'' quantities give rise to the s-e tensors.
In the full Einstein-Maxwell theory, there exists a divergence-free 4-index
tensor \cite{PR} which, unfortunately, does not satisfy DSEP nor (\ref{cero})
and, due to its peculiar index symmetries,
it does not provide conserved currents.

To explore the possibility of s-e exchange, let us consider the propagation of
discontinuities \cite{S}. Assume there is an electromagnetic field propagating
in a background so that there is a (necessarily null) hypersurface of
discontinuity $\Sigma$. Let us denote by $[V]_{\Sigma}$ the discontinuity of any
object $V$ across $\Sigma$. From the classical Hadamard theory one proves the
existence of $c_{\mu}$ on $\Sigma$ such that
\ben
\left[F_{\mu\nu}\right]_{\Sigma}=n_{\mu}c_{\nu}-n_{\nu}c_{\mu}, \hspace{5mm}
n_{\rho}c^{\rho}=0
\een
where $n_{\mu}$ is a null 1-form normal to $\Sigma$ (notice that ${\vec n}$
is in fact tangent to $\Sigma$ because $n_{\mu}n^{\mu}=0$ \cite{MS}), as well
as the following propagation equation 
\ben
n^{\mu}\partial_{\mu}c^2+c^2(K+2\psi)=0,\hspace{5mm}
c^2\equiv c_{\mu}c^{\mu}\geq 0
\een
where $\psi$ is the factor appearing in $n^{\mu}\nabla_{\mu}n^{\nu}=\psi n^{\nu}$
and $K$ is the trace of the second fundamental form of $\Sigma$ relative
to $n_{\mu}$ (as $\Sigma$ is null, such a second fundamental form is orthogonal
to ${\vec n}$ and not extrinsic \cite{MS}). The above
propagation equation ensures that if $c_{\mu}=0$ at a spacelike 2-surface cut
$S$ of $\Sigma$, then $c_{\mu}=0$ everywhere on $\Sigma$. For any conformal
Killing vector ${\vec \zeta}$ it is then straightforward to prove that
\ben
\int_{S} c^2 \left(n_{\mu}\zeta^{\mu}\right)^2 \bm{\omega}
\een
is a conserved quantity along $\Sigma$, where $\bm{\omega}$ is the volume
element 2-form of $S$. This can be related to the energy-momentum of the
electromagnetic field because
$\left[T_{\mu\nu}\{F_{[2]}\}\right]_{\Sigma}=c^2 n_{\mu}n_{\nu}$
if the background is empty.

But what happens when $\left[F_{\mu\nu}\right]_{\Sigma}=0$? Then, 
$T_{\mu\nu}\{F_{[2]}\}$ is continuous and s-e quantities must come in.
Now there exist $B_{\mu\nu}$ and $f_{\mu}$ on $\Sigma$ such that
\ben
\left[R_{\alpha\beta\lambda\mu}\right]_{\Sigma}=
4n_{[\alpha}B_{\beta][\mu}n_{\lambda]}, \hspace{5mm} B_{\mu\nu}=B_{\nu\mu}
\hspace{5mm} 2n^{\mu}B_{\mu\beta}+g^{\mu\nu}B_{\mu\nu}n_{\beta}=0, \\
\left[\nabla_{\alpha}F_{\mu\nu}\right]_{\Sigma}=2 n_{\alpha} n_{[\mu}f_{\nu]},
\hspace{5mm} n_{\rho}f^{\rho}=0 \hspace{25mm}
\een
and satisfying the propagation laws \cite{L}
\bea
n^{\mu}\partial_{\mu}B^2+B^2(K+4\psi)-
2n^{\sigma}F_{\sigma\rho}B^{\rho\tau}f_{\tau}=0, \hspace{5mm} 
B^2\equiv B_{\mu\nu}B^{\mu\nu}\geq 0, \label{1}\\
n^{\mu}\partial_{\mu}f^2+f^2(K+4\psi)+
2n^{\sigma}F_{\sigma\rho}B^{\rho\tau}f_{\tau}=0, \hspace{5mm}
f^2\equiv f_{\mu}f^{\mu}\geq 0 \, .\label{2}
\eea
Therefore, the discontinuities of the Riemann tensor and of
$\nabla_{\alpha}F_{\mu\nu}$ are a source for each other. In fact, from
(\ref{1}) and (\ref{2}) it is immediate that \cite{L}
\ben
n^{\mu}\partial_{\mu}\left(B^2+f^2\right)+\left(B^2+f^2\right)(K+4\psi)=0
\een
from where one finds the following conserved quantity along $\Sigma$
\be
\fbox{$\displaystyle{\int_{S} \left(B^2+f^2\right)
\left(n_{\mu}\zeta^{\mu}\right)^4 \bm{\omega}}$} \label{3}
\ee
which needs {\it both} the electromagnetic and gravitational contributions.
The interesting point is that (\ref{3}) is related to the
tensors (\ref{bel}) and (\ref{em}): if $\nabla_{\alpha}F_{\mu\nu}$ vanishes at
one side of $\Sigma$, then $\left[B_{\alpha\beta\mu\nu}\right]_{\Sigma}=
2B^2 n_{\alpha}n_{\beta}n_{\mu}n_{\nu}$ and 
$\left[T_{\alpha\beta\mu\nu}\{\nabla_{[1]}F_{[2]}\}\right]_{\Sigma}=
2f^2 n_{\alpha}n_{\beta}n_{\mu}n_{\nu}$.

To check whether or not this interplay of s-e quantities is generic, 
consider finally a scalar field $\phi$ satisfying the Klein-Gordon equation
$\nabla_{\rho}\nabla^{\rho}\phi =m^2\phi$ for mass $m$. Expression (\ref{set})
for $\phi$ would simply be $T\{\phi\}=\phi^2/2$.
However, the case of physical interest arises by constructing the tensor
(\ref{set}) associated to $\nabla _{\mu}\phi$:
\ben
T_{\lambda\mu}\{\nabla_{[1]}\phi\}=\nabla_{\lambda}\phi\nabla_{\mu}\phi
-\frac{1}{2}g_{\lambda\mu}\nabla_{\rho}\phi\nabla^{\rho}\phi
\een
which is the standard energy-momentum tensor of a {\it massless} scalar field.
When $m\neq 0$, $T\{m\phi\}$ has the same physical dimensions than
$T_{\lambda\mu}\{\nabla_{[1]}\phi\}$, so that one must combine them (using
$-g_{\mu\nu}$ which obviously satisfies DSEP) to obtain the whole
energy-momentum tensor of the minimally coupled scalar field
\ben
T_{\lambda\mu}\equiv T_{\lambda\mu}\{\nabla_{[1]}\phi\}+T\{m\phi\}
(-g_{\lambda\mu})=\nabla_{\lambda}\phi\nabla_{\mu}\phi
-\frac{1}{2}g_{\lambda\mu}\nabla_{\rho}\phi\nabla^{\rho}\phi -
\frac{1}{2}m^2\phi^2 g_{\lambda\mu}\, .
\een
This procedure is systematic and the BR-like tensor for the scalar field is
the sum of expression (\ref{set}) for the double symmetric (1,1)-form
$\nabla_{\alpha}\nabla_{\beta}\phi$ plus the terms coming from $m$.
Expanding the duals one gets  
\bea
T_{\alpha\beta\lambda\mu}\!\equiv
T_{\alpha\beta\lambda\mu}\{\nabla_{[1]}\nabla_{[1]}\phi\}-\!
T_{\alpha\beta}\{m\nabla_{[1]}\phi\}g_{\lambda\mu}\!-\!
T_{\lambda\mu}\{m\nabla_{[1]}\phi\}g_{\alpha\beta}+\!
T\{m^2\phi\}g_{\alpha\beta}g_{\lambda\mu}\!= \nonumber \\
=2\nabla_{\alpha}\nabla_{(\lambda}\phi \nabla_{\mu)}\nabla_{\beta}\phi
-g_{\alpha\beta}\left(\nabla_{\lambda}\nabla^{\rho}\phi
\nabla_{\mu}\nabla_{\rho}\phi+m^2\nabla_{\lambda}\phi\nabla_{\mu}\phi \right)
- g_{\lambda\mu}\left(\nabla_{\alpha}\nabla^{\rho}\phi
\nabla_{\beta}\nabla_{\rho}\phi \right. \nonumber \\
\left. +m^2\nabla_{\alpha}\phi\nabla_{\beta}\phi\right)
+\frac{1}{2} g_{\alpha\beta}g_{\lambda\mu}\left(
\nabla_{\sigma}\nabla_{\rho}\phi\nabla^{\sigma}\nabla^{\rho}\phi +
2m^2\nabla_{\rho}\phi\nabla^{\rho}\phi +m^4\phi^2\right)
\label{sc} \hspace{1cm}
\eea
which satisfies DSEP,
$T_{\alpha\beta\lambda\mu}=T_{(\alpha\beta)(\lambda\mu)}=
T_{\lambda\mu\alpha\beta}$, and was previously found in Special Relativity
\cite{BT}. Again one can construct (super)$^n$-energy tensors associated to the
higher derivatives $\nabla_{\mu_1}\dots\nabla_{\mu_{n+1}} \phi$:
{\it the (super)$^n$-energy
(tensor) of the scalar field ($m\neq 0$) vanishes at a point $p$ iff the
covariant derivatives of $\phi$ up to (n+1)$^{th}$-order are zero at $p$}.
(If $m=0$ the result involves the (n+1)$^{th}$-derivatives exclusively).

Direct computation leads to
\bea
\nabla_{\alpha}T^{\alpha}_{\hspace{1mm}\beta\lambda\mu}=
2\nabla_{\beta}\nabla_{(\lambda}\phi
R_{\mu)\rho}\nabla^{\rho}\phi -g_{\lambda\mu} R^{\sigma\rho}
\nabla_{\beta}\nabla_{\rho}\phi\nabla_{\sigma}\phi \nonumber\\
-\nabla_{\sigma}\phi \left(2\nabla^{\rho}\nabla_{(\lambda}\phi \,
R^{\sigma}_{\mu)\rho\beta} +g_{\lambda\mu}
R^{\sigma}_{\rho\beta\tau}\nabla^{\rho}\nabla^{\tau}\phi\right) \label{div2}
\eea
and similar longer formulae for (super)$^n$-energy tensors, so that they
are {\it divergence-free in flat spacetime}, providing infinitely many
conserved quantities there.
\newpage

The situation hitherto is that the Bel tensor (\ref{bel}) is
divergence-free in vacuum, and the s-e tensor (\ref{sc}) of the scalar field
is divergence-free in the absence of gravitation. The natural question arises:
can these tensors be combined to produce a conserved quantity in General
Relativity? To answer it, assume that the spacetime satisfies the
Einstein-Klein-Gordon equations, so that
\be
R_{\mu\nu}=\nabla_{\mu}\phi\nabla_{\nu}\phi +\frac{1}{2}m^2\phi^2g_{\mu\nu}
\label{ric}
\ee
and that ${\vec \xi}$ is a Killing vector. Then $\xi^{\mu}\nabla_{\mu}\phi =0$
(if $m\neq 0$) \cite{Pa} from where it also follows
$\xi^{\mu}\nabla^{\nu}\phi\nabla_{\mu}\nabla_{\nu}\phi =0$.
Using these results with (\ref{div}) and (\ref{ric}) one finds
\ben
\nabla_{\alpha}\left(B^{\alpha\beta\lambda\mu}\xi_{\beta}\xi_{\lambda}\xi_{\mu}
\right)=\nabla_{\sigma}\phi 
\left(2\nabla^{\rho}\nabla_{\lambda}\phi
R^{\sigma}_{\mu\rho\beta} +g_{\lambda\mu}R^{\sigma}_{\rho\beta\tau}
\nabla^{\rho}\nabla^{\tau}\phi\right)\xi^{\beta}\xi^{\lambda}\xi^{\mu}
\een
and from (\ref{div2}) and (\ref{ric})
\ben
\nabla_{\alpha}\left(T^{\alpha\beta\lambda\mu}\xi_{\beta}\xi_{\lambda}\xi_{\mu}
\right)=-\nabla_{\sigma}\phi 
\left(2\nabla^{\rho}\nabla_{\lambda}\phi
R^{\sigma}_{\mu\rho\beta} +g_{\lambda\mu}R^{\sigma}_{\rho\beta\tau}
\nabla^{\rho}\nabla^{\tau}\phi\right)\xi^{\beta}\xi^{\lambda}\xi^{\mu}
\een
so that finally
\ben
\fbox{$\nabla_{\alpha}j^{\alpha}=0$} \hspace{1cm}
j^{\alpha}\equiv \left(B^{\alpha\beta\lambda\mu}+
T^{\alpha\beta\lambda\mu}\right)\xi_{\beta}\xi_{\lambda}\xi_{\mu}\, .
\een
This also holds when $m=0$ for either case of ${\vec \xi}$ being
parallel or orthogonal to $\nabla \phi$. The relevance of this result is that
provides {\it conserved s-e quantities} (by means of the typical
integration of $j^{\alpha}$ and Gauss' theorem) {\it proving the interchange
of s-e between the gravitational and scalar fields}, because neither
$B^{\alpha\beta\lambda\mu}\xi_{\beta}\xi_{\lambda}\xi_{\mu}$ nor
$T^{\alpha\beta\lambda\mu}\xi_{\beta}\xi_{\lambda}\xi_{\mu}$ are divergence-free
separately in general. 
These conservation laws make definition (\ref{set}) of s-e tensors not only
mathematically appealing but also physically very promising.

\section*{Acknowledgements}
I am grateful to Llu\'{\i}s Bel, G\"oran Bergqvist, Marc Mars, Pierre
Teyssandier and Ra\"ul Vera for their help and comments.

\end{document}